# Broadband transverse susceptibility in multiferroic Y-type hexaferrite $Ba_{0.5}Sr_{1.5}Zn_2Fe_{12}O_{22}$

P. Hernández-Gómez*, D. Martín-González, C. Torres, J. M. Muñoz

Dpt. Electricidad y Electrónica, Universidad de Valladolid, Paseo de Belén 7, 47011 Valladolid Spain.

*Corresponding author

e-mail: pabloher@ee.uva.es

Tel: +34983423895

**ABSTRACT**

Noncollinear spin systems with magnetically induced ferroelectricity from changes in spiral magnetic ordering have attracted significant interest in recent research due to their remarkable magnetoelectric effects with promising applications. Single phase multiferroics are of great interest for these new multifunctional devices, being Y-type hexaferrites good candidates, and among them the ZnY compounds due to their ordered magnetic behaviour over room temperature. Polycrystalline Y type hexaferrites with composition $Ba_{0.5}Sr_{1.5}Zn_2Fe_2O_{22}$ (BSZFO) were sintered in 1050º C-1250º C temperature range. Transverse susceptibility measurements carried out on these BSZFO samples in the temperature range 80-350 K with DC fields up to ±5000 Oe reveal different behaviour depending on the sintering temperature. Sample sintered at 1250 ºC is qualitatively different, suggesting a mixed Y and Z phase like CoY hexaferrites. Sintering at lower temperatures produce single phase Y-type, but the transverse susceptibility behaviour of the sample sintered at 1150 ºC is shifted at temperatures 15 K higher. Regarding the DC field sweeps the observed behaviour is a peak that shifts to lower values with increasing temperature, and the samples corresponding to single Y phase exhibit several maxima and minima in the 250 K – 330 K range at low DC applied field as a result of the magnetic field induced spin transitions in this compound.

## Introduction

Magnetoelectric multiferroics have attracted significant interest in recent research activities because they can present unusual physical phenomena like remarkable magnetoelectric effects with potential applications in ultra-dense magnetic storage devices, as low power spintronic devices, or effective medical drug delivery [1-4]. Among the compounds with this characteristic, ferroxplana type hexaferrites with Z and Y phases are very promising materials, due to the giant magnetoelectric coupling between magnetism and ferroelectricity caused by the spin transitions from longitudinal to transverse helimagnetic phases and the low magnetic fields to switch the electric polarization [5-7]. Particular interest is devoted to Y type structures with composition ($[BaSr]_2Me_2Fe_{12}O_{22}$) because pure phase formation of Z type is challenging.

The crystal structure of Y-type hexaferrite belongs to the R-3m space group, obtained with the piling up of S ($Me_2Fe_4O_8$) and T ($BaFe_8O_{14}$) blocks along the c axis. Metal cations occupy four octahedral ($18h_{VI}$, $6c_{VI}$, $3a_{VI}$ and $3b_{VI}$) and two tetrahedral sites ($6c_{IV}$ and $6c_{IV}^*$) [20]. In the last years several compounds belonging to this structural phase have been identified as good candidates regarding their magnetoelectric properties, namely $Ba_2Mg_2Fe_{12}O_{22}$ [8-10], $Ba_{0.5}Sr_{1.5}Co_2Fe_{12}O_{22}$ (BSFCO) [11, 12], $Ba_{0.5}Sr_{1.5}Zn_2Fe_{12}O_{22}$ (BSZFO) [13, 14] and Al substituted variants $Ba_{0.5}Sr_{1.5}Co_2(FeAl)_{12}O_{22}$ (BSCFAO) [15-17] and $Ba_{0.5}Sr_{1.5}Zn_2(FeAl)_{12}O_{22}$. (BSZFAO) [18, 19].

The magnetic structure is ferrimagnetic, with the layers containing Ba and Sr ions having small net magnetic moments, whereas the rest of layers have large magnetic moments, so that it has been described as S and L blocks [2], with a net moment perpendicular to the c axis. The progressive replacement of Ba by Sr lead to a lattice deformation that increase the superexchange bond angles, in particular those corresponding to the metal cations in the layers of Ba(Sr) ions. As a consequence of this, the compound can exhibit different spiral magnetic structures: proper screw, longitudinal conical (LC), transverse conical (TC), intermediates or collinear [1, 8], some of which allow the existence of an induced electric polarization that can be explained according to the inverse Dzyaloshinskii-Moriya effect or spin current model [1]. Recent works also point to the existence of spin chirality and chiral domains in these compounds [17, 21, 22]

Transverse susceptibility (TS) is obtained when the magnetic response of a material is measured in the direction of a small AC applied field perpendicular to a bias DC magnetic field. Is a versatile tool to study singular properties of magnetic systems, especially to obtain their anisotropy and switching field [23, 24], and is also a probe of phase transitions caused by anisotropy [25]. Despite the early description of this kind of measurement [26] and its application [23], it is still of interest in nanoparticle magnetic systems [27, 28]. In this work TS measurements have been carried out with a broadband system based on a LCR meter [29] on Y type hexaferrites with composition $Ba_{0.5}Sr_{1.5}Zn_2Fe_{12}O_{22}$, (BSZFO) optimal to exhibit multiferroic properties [14]. According with our previous experience we expect that the different spiral magnetic configurations alter the TS results.

**Experimental**

Polycrystalline samples have been prepared by means of standard ceramic techniques. Stoichiometric amounts of $SrCO_3$ (98%), $BaCO_3$ (99%), ZnO (99%), and $Fe_2O_3$ (99%) with molar proportions 4.5: 1.5: 2: 18, corresponding to the initial composition of Y type hexaferrite $Ba_{0.5}Sr_{1.5}Zn_2Fe_{12}O_{22}$ (BSFZO), were mixed in agate mortar, calcined at 900 ºC to remove carbonates, pressed in the form of rods of 5 mm diameter and 20 mm length, and then sintered in air atmosphere at 1050 °C, 1150 °C and 1250 °C (samples A, B and C respectively), according to the phase diagrams [20].

X ray diffraction patterns were obtained with a diffractometer Bruker Discover 8 (LTI lab at Valladolid University) at room temperature employing Cu-K$\alpha$ radiation ($\lambda$=1.54056 Å). Intensity data were collected by the step-counting method (step 0.02 °/s) in the range 20° < 2$\theta$ <70°. Quasi static hysteresis loops were obtained in powdered samples with an inductive technique at room temperature with a maximum field of 7500 Oe. Homemade control program also corrects for the skew of the loop due to shape demagnetization factor.

Transverse susceptibility measurements were carried out with a broadband automatic system [29] in which the sample rods form the core of a coil that produces a longitudinal AC magnetic field with frequency 1 kHz and maximum amplitude of 2 Oe. The sampleholder with the coil and a heating element is put inside a cryostat to allow temperature measurements in the range 80 K- 350 K. The cryostat tail lies into the polar pieces of an electromagnet fed by a power supply Agilent 6675A that produces a DC magnetic field, measured with a FW Bell 5080 gaussmeter, perpendicular to the AC magnetic field. DC bias field sweeps run from +5000 Oe to -5000 Oe, and then sweep back to +5000 Oe (i.e. bipolar scans), according to measurements found in literature [30]. The response of the measuring coil is obtained with a precision LCR meter Agilent E4980A. Temperature control is achieved with a data logger Hewlett Packard 3497A. All the system is controlled via GPIB with a PC by means of a homemade control program with Agilent VEE software. The broadband nature of our system arises from the possibility of varying temperature, DC magnetic field, AC field frequency and amplitude with enhanced sensitivity.

**Results**

In Figure 1, the XRD patterns of sintered Y type hexaferrite samples are presented together with the pattern corresponding to the JCPDS cards 40-1047 and 73-2036, corresponding to Y and Z type structures. The diffraction patterns reveal that samples sintered at 1050 °C and 1150 °C (samples A and B) are single phase Y-type hexaferrites with space group R-3m [11]. Sample C corresponds to Y type hexaferrite with secondary phase of Z type, together with $SrFe_2O_4$, so that the sintering temperature effect on phase formation is very similar to the observed in previous works [29]. The lattice parameters and particle sizes obtained for these samples reveal minor changes in the Y-type phase with the increase in sintering temperature: a= 5.8689 Å, and c= 43.4760 Å for the sample A, and 5.8634 Å, and 43.4468 Å for the sample B, whereas crystallite sizes are both around 100 nm (from Scherrer formula) and particle sizes of 2-6 μm (from TEM results, not shown) in good agreement with literature data [13, 31, 32].

Hysteresis loops of the sintered Y type hexaferrites obtained at room temperature (295 K) are shown in Figure 2. We can observe very different loop shapes with the different sintering temperatures analysed. Loops are rather narrow, according to the soft magnetic character of ferroxplana compounds. At the higher sintering temperature tested higher crystallinity and densification is obtained, resulting in higher magnetization and smaller coercive fields, caused by the higher crystallinity [33], the increased contribution of the secondary Z phase, as well as the absence of magnetic phase transitions near room temperature in which this measurement has been carried out. With lower sintering temperatures we have pure Y type hexaferrite, however, the hysteresis loops are very different: in sample B the loop is narrow and some bends can be observed at 300 and 1000 Oe, as previously reported by other authors [32, 34], probably caused by the undergoing field induced magnetic transitions that also take place in polycrystals [32]. This bends are also observed in minor loops (not shown), revealing that the magnetic phase transitions take place with a rather low field. On the other hand, in sample A the loop is wider than the others and bends are not observed, unless it is stretched near zero field. The coercive fields are 231.2 Oe, 111.3 Oe and 42.6 Oe, and the remanent magnetization are 2.31 emu/g, 0.51 emu/g and 1.75 emu/g for A, B and C samples, respectively. The use of the law of approach to saturation can give us a reference value of saturation magnetization in these compounds: 12.8 emu/g, 13.0 emu/g and 27.5 emu/g for A, B and C samples (data have been converted to cgs units to allow comparison with literature data). We can observe a clear increase in saturation magnetization in sample C, caused by the secondary phase with absence of spin transitions near room temperature in which the hysteresis temperature measurements have been carried out.

The transverse susceptibility ratio can be expressed as [28, 30]

$$\frac{\Delta \chi_T}{\chi_T}(\%) = \frac{\chi_T(H) - \chi_T(H_{SAT})}{\chi_T(H_{SAT})} \times 100 \qquad (1)$$

where $\chi_T(H_{SAT})$ is the transverse susceptibility at the saturating field ($H_{SAT}$ =5000 Oe in our case). For samples filling completely the measuring coil we can deduce [28]

$$\frac{\Delta \chi_T}{\chi_T}(\%) \approx \frac{L(H) - L(H_{SAT})}{L(H_{SAT})} \times 100 \qquad (2)$$

where L(H) is the inductance of the test coil with the sample at the different DC applied fields. Since this is a measure of the overall change in transverse susceptibility, there is no dependence on geometrical parameters, and then it is useful to extract several parameters, such transition temperatures, anisotropy and switching fields in many systems [25, 27, 28, 30]. When $\Delta\chi_T/\chi_T$ is represented as a function of DC field, maxima are observed at $H_{DC}=\pm H_A$ and the effective anisotropy constant can be deduced. The theoretically expected maximum at the switching field $H_{DC}=H_s$. is often merged to one of the former in systems with distribution of anisotropy fields.

In the figure 3 the 3D plots of the bipolar scans of broadband TS measurements obtained in the conditions expressed in the previous section are shown. The repeatability of the TS results has been checked by repeating the measurements, starting at different temperatures, and employing a bias DC field of half the maximum value reported here, obtaining the same results. The common trend is the gradual diminution of TS maximum value with the increase in sintering temperature: from 25% for the sample A, to 20% and 14% for samples B and C respectively. The main contribution to the TS come from the grains whose easy axis is oriented perpendicular to the DC field [23], so that higher crystallite size reduce the amount of grains with the optimal orientation to the TS signal, thus lowering its amplitude. However, we can see clear differences among the sample sintered at 1250 ºC and the others. In the former we can observe that at low measuring temperatures the amplitude of TS is small and increases strongly over room temperature, being this behaviour similar to the Y-type system prepared with Co [29]. On the contrary, the latter exhibit a wide region at low temperatures with TS amplitude close to the maximum, a sudden decrease around room temperature, and a further increase to a lower TS peak around 7 % to finally reduce to zero when the temperature goes over the Nèel temperature. The main difference among these two samples is that in sample B the temperatures in which the above mentioned increases and decreases of the TS values take place are 15 K higher than in sample A, as well as the Néel temperatures, that are 320 K and 335 K for samples A and B respectively.

In the figure 4, representative 2D plots of the TS measurements depending of the DC applied magnetic field are shown. We can observe two common behaviours in all the samples with the measuring temperature($T_{meas}$): at higher $T_{meas}$ we can observe only one peak, while at low $T_{meas}$ it can be seen that a bipolar scan starting from $+H_{SAT}$ exhibits a maximum at $+H_A$, and in the sweep from 0 to $-H_{SAT}$ a maximum with a shoulder is observed. The sweep from negative to positive $H_{SAT}$ is similar. With this representation we obtain that the anisotropy values decrease from 300 Oe to 80 Oe in sample A, and from 500 Oe to 80 Oe in sample B, when $T_{meas}$ is increased from 80 to 250 K. The TS behaviour is strongly dependent on the sintering temperature, and we can observe different qualitative responses for the sample C regarding the others. In this sample the peak related to anisotropy at low $T_{meas}$ shifts slightly to higher applied fields, from 300 Oe at $T_{meas}$ =80 K to 400 Oe at $T_{meas}$ =220 K, and then reduces to zero at the top measuring temperature tested. This behaviour is to some extent similar to that observed in CoY hexaferrites [29], unless the comparative analysis of these two systems will be postponed to a further paper.

On the other hand, the TS behaviour of the samples sintered at lower temperatures has a striking feature in the region around room temperature not observed before in any magnetic system. The TS measurements of the sample B in the temperature range 260 K – 320 K are represented in figure 5 (for clarity, we only represent the unipolar sweep from +Hsat to –Hsat, as both unipolar sweeps are identical). We can see that the TS amplitude decreases suddenly at temperatures over 260 K, whereas at the same time the TS field profile changes as follows:

- In the region 260 K- 280K the curve changes from a single maximum in the positive field region, that shifts to a value almost zero with increasing $T_{meas}$, to a curve with this maximum together with a minimum and a maximum in the negative field region.

- Over 280 K and up to 305 K, an additional maximum and an additional minimum are also observed in the positive region, at fields around 1000 Oe and 300 Oe respectively. The DC magnetic field required for obtaining this secondary maximum shifts to higher values with $T_{meas}$ whereas the minimum remains almost in a constant value of DC field.

-Then at 310 K we observe the onset of a different behaviour with higher TS amplitude, the minimum disappear and maxima become symmetrical and move back to lower DC field values.

- Finally at 320 K and up to the Neel temperature at 335 K, the single TS peak is located at zero applied field.

It is worth mention that in sample A the behaviour is very similar but the temperature ranges in which the TS variations occur are 15 K lower, as mentioned previously.

We could consider if the type of measurement done or their effects on the different grains of the polycrystalline sample could distort the results, but this possibility has to be ruled out. If the DC bias field is parallel to the easy axis of the ferrite the alternating magnetic field does not have influence until the DC field approaches to zero. Then some spin canting is favoured, promoting a conical magnetization with very low angle regarding the easy plane. For grains with c axis parallel to DC field, normal or alternate longitudinal conical magnetization is favoured. Alternating field will have more effect as it is aligned with the easy axis, and when bias field diminishes it will promote conical orientation. Anyway, this effect is really small taking into account the intensity of alternating field regarding the DC fields in which spin transitions takes place, and their negligible contribution has been proved by increasing the AC field one order of magnitude without noticeable variations in the TS profile.

**Discussion**

According to previous studies [7, 14], the BSZFO system undergoes an antiferromagnetic transition at 337 K after a sudden decrease of magnetization over 315 K, that is connected with a metamagnetic spin transition [18]. This transition is related with the sudden change of the angle from 134º to 180º (i.e. the collinear planar structure) [13]. It is also known that these transition temperatures can be reduced up to 19 K due to sintering conditions [18, 34]. In particular, the results reported by Kamba et al [34] with magnetization measurements attribute to the cooling stage in sample fabrication this temperature variation. As all our samples are produced with identical sintering profiles our results are noteworthy, because they reveal that not only the slow cooling but the top sintering temperature plays a critical role. Taken into account that we have the same crystalline phase in samples A and B, minor changes in particle size due to the different sintering temperature and perhaps some different distribution of metallic cations, especially the Zn cation, that can attain different ordering in quenched or in slow cooled ferrites, are responsible for the 15 K temperature shift observed for the Nèel as well as the magnetic field induced spin transitions. As the hysteresis measuring temperature is kept constant at 295 K in our experiment, this shift causes that samples A and B are at different spin states at this temperature, explaining why we obtained different loops, and the loop shapes reported here can be an indicative tool of the presence of spin transitions.

In addition, under an applied magnetic field different magnetization curves have been obtained at different measuring temperatures, some of them with bends caused by the undergoing spin transitions that take place in BSZFO. The different magnetic structures change from planar helix to collinear ferrimagnetic ordering, with several intermediate commensurate magnetic spin phases with mixed conical structure, called I, II and III (the spin orientations in this phases can be seen in [4, 8, 19]). For each composition, some of the above, as well as alternate longitudinal conical (ALC) and transverse conical (TC) can take place at different values of temperature and applied magnetic field leading to a complex magnetic phase diagram [4, 8, 35, 36]. Unless the spin induced ferroelectric order is usually attributed to the TC spiral system, other authors have suggested that the intermediate II and III spin systems can also promote electric polarization [1, 4, 37]. Also, the existence of spin-chiral domains[17] or slainted conical spin structures in the ALC phase [4] can cause strong polarization and hence multiferroic behaviour [15]. From the results obtained with the transverse susceptibility and hysteresis loops, applied fields around 300 and 1000 Oe are unambiguously related to two field induced spin transitions that take place in the ZnY hexaferrite in the range 280- 305 K for the sample B, or 265-290 K for the sample A. Our measurement setup is then a sensitive tool to determine the applied field required to promote the spin transitions at each $T_{meas}$. According to the magnetic phase diagrams [4, 35, 36] the spin transitions involved can be planar helix, intermediate II and intermediate III with increasing field.

It is noteworthy that in magnetodielectric studies the effect of magnetic field on dielectric permittivity have similar temperature range in which the peak moves to zero and finally disappears, showing close relationship among the transverse susceptibility and the magnetodielectric effect measurements [38, 39]. In addition, the fabrication conditions have a critical effect on the temperature and magnetic field ranges in which the magnetoelectric effects take place [38]. It has been suggested that the ferroelectric phase not only becomes independent of the existence of intermediate III phase but that it is a generic property of this hexaferrite system [38], and that with some substitutions this compound exhibits a heliconical spin ground state or at least a mixed helical and heliconical spin state able to induce electrical polarization at temperatures between 280 and 315 K and under fields lower than 1000 Oe [37]. With the increase of temperature over 280 K, the in-plane anisotropy that confine the spins within the easy plane weakens, so that the conical structure appear at lower fields [13], in good agreement with our TS results in which the secondary maximum shifts to lower fields with increasing the measuring temperature (see figure 5).

We have observed that spin transition temperatures increase from sample A to sample B without any structural phase change and with minor lattice variation. This can be due to the different amount of Zn cations in the tetrahedral sites in the L- S block interface layer [32], indicating that the sintering conditions plays a critical role on the behaviour of the interaction among L and S blocks that define the magnetization. Some occupation of Zn cations or vacancies in octahedral sites could act in a similar way than non magnetic dopants like Al, increasing the anisotropy along c axis and promoting the heliconic magnetization [36]. Off centered distortion in the cation placement induce the existence of a non negligible out of plane magnetic orbital moment that also contribute to the presence of helical magnetization with planar applied fields [40]. The reduction of the in-plane anisotropy can promote heliconical magnetization that is revealed by different magnetization curves parallel and perpendicular to c axis in the range from -1000 Oe to +1000 Oe [40]. As we have observed anomalies in this field range, we can state that we are facing a similar process. In our case the anomalies are caused by effect of the increased sintering temperature in the higher amount of cation vacancies acting in a similar way to non magnetic substitutions, i.e, promoting spin transitions with increasing field, in which a mixed conical state allow the existence of peaks in magnetodielectric effects and variations in transverse magnetic susceptibility, so that low field magnetoelectric effects can be obtained near room temperature.

**Conclusions**

Transverse susceptibility measurements have been carried out on BSZFO polycrystalline samples sintered in 1050º C-1250º C temperature range, revealing different behaviour depending on the sintering temperature. The relative amplitude of TS decreases with the increase in sintering temperature. Sample sintered at 1250 ºC is qualitatively different, suggesting a mixed Y and Z phase like. By sintering at lower temperatures single phase Y-type compounds are obtained, but the TS behaviour of the sample sintered at 1150 ºC is shifted at temperatures 15 K higher, revealing that sintering conditions play a key role in the properties. Regarding the DC field sweeps the observed behaviour is a peak that shifts to lower values with increasing temperature. However, the samples corresponding to single Y phase exhibit several maxima and minima in the 250 K – 330 K range at low DC applied field as a result of the undergoing magnetic phase transitions in the BSZFO system, involving mixed conical states. The heliconical spin state that appears between 280 and 315 K could generate electric polarization under fields lower than 1000 Oe, due to the reduction in the spin anisotropy, improving low field magnetoelectric response and allowing multiferroic behaviour. Broadband TS measurements thus provide a valuable information about the temperature and magnetic field ranges of the spin transitions in these compounds.

**Acknowledgements**

Funding: This work was supported by the Spanish Ministerio de Ciencia Innovación y Universidades, (AEI with FEDER), project id. MAT2016-80784-P. Assistance of Mr. M. Suazo in sample fabrication is acknowledged. D. Martín-González acknowledges "Junta de Castilla y León, Estrategia de Emprendimiento y Empleo Joven, Fondo Social Europeo" contract.

**Figure Captions**

**Figure 1**. X ray diffractograms of the sintered $Ba_{0.5}Sr_{1.5}Zn_2Fe_{12}O_{22}$ hexaferrite sample at 1050º C, 1150º C and 1250º C (samples A, B and C respectively), and the corresponding JCPDS cards.

**Figure 2**. Hysteresis loops of the $Ba_{0.5}Sr_{1.5}Zn_2Fe_{12}O_{22}$ hexaferrite samples sintered at 1050º C, 1150º C, and 1250º C (samples A, B and C respectively). In the inset the behaviour in the low field region is magnified.

**Figure 3**. 3D view of the broadband transverse susceptibility of the $Ba_{0.5}Sr_{1.5}Zn_2Fe_{12}O_{22}$ hexaferrite samples sintered at a) 1050 ºC b) 1150 ºC and c) 1250 ºC ( samples A, B and C respectively).

**Figure 4**. 2D plot of transverse susceptibility vs DC magnetic field of the $Ba_{0.5}Sr_{1.5}Zn_2Fe_{12}O_{22}$ hexaferrite samples sintered at a) 1050 ºC b) 1150 ºC and c) 1250 ºC. (samples A, B and C respectively).

**Figure 5**. Near room temperature behaviour of TS measurements in $Ba_{0.5}Sr_{1.5}Zn_2Fe_{12}O_{22}$ hexaferrite sample sintered at 1150 ºC (sample B).

Figure 1

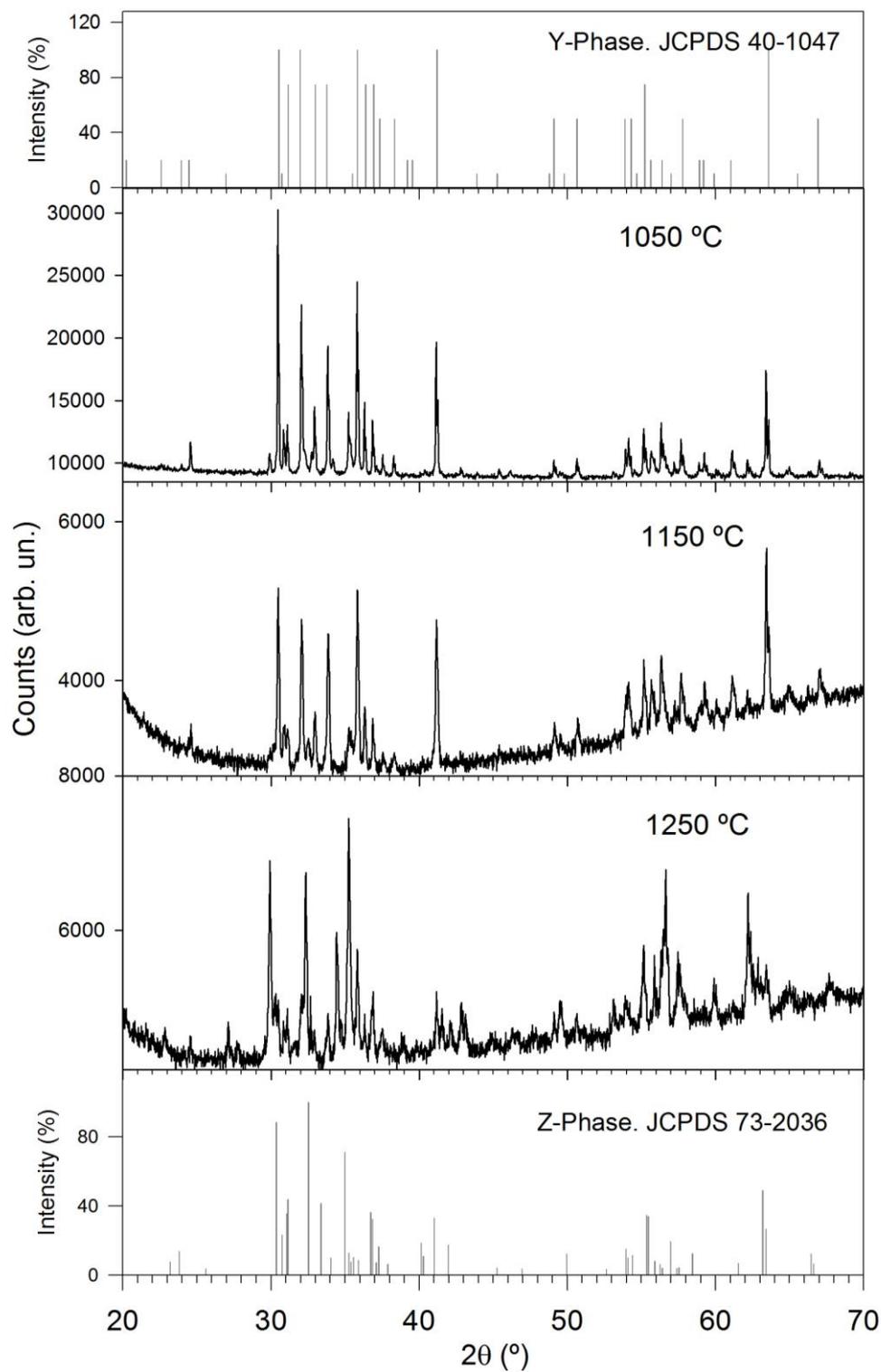

Figure 2

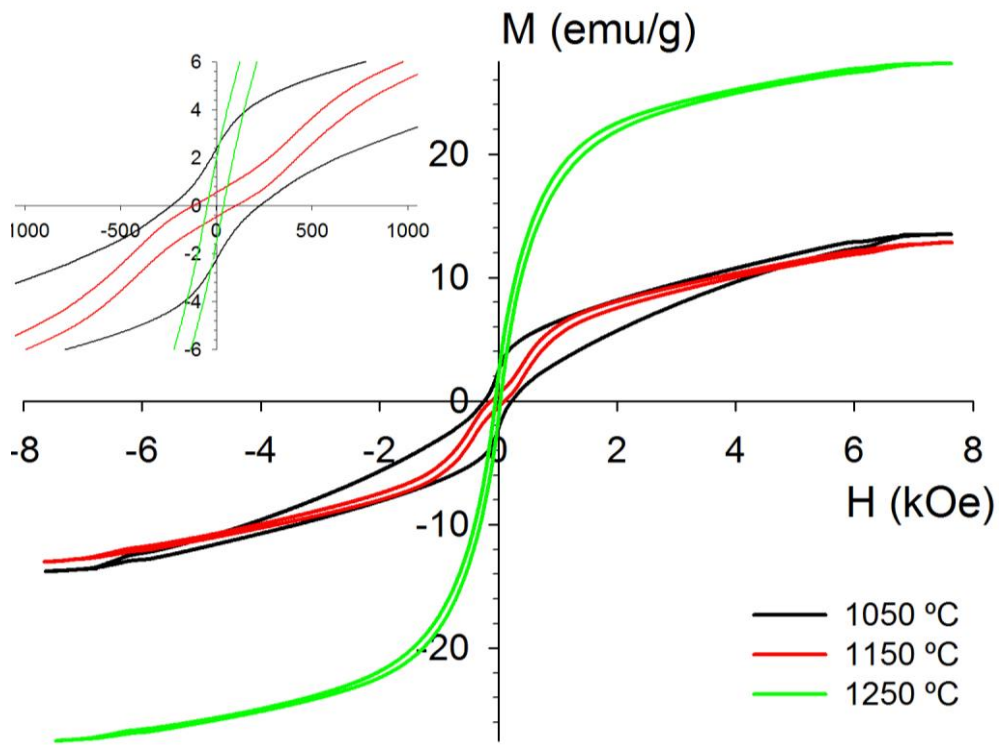

Figure 3

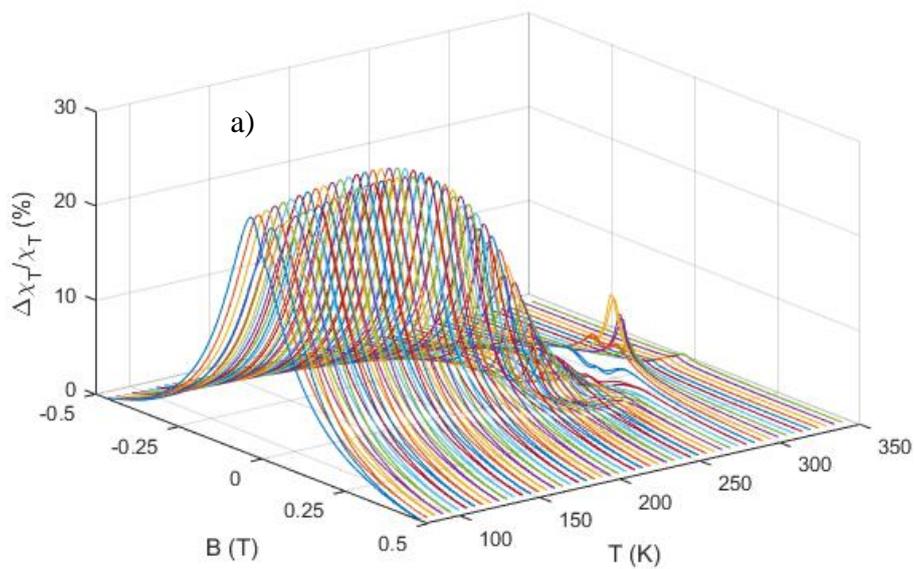

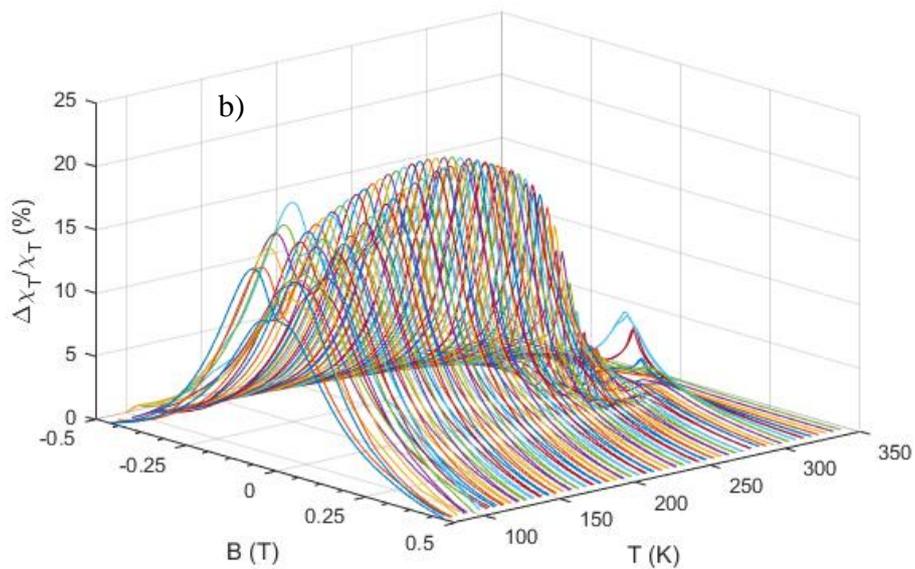

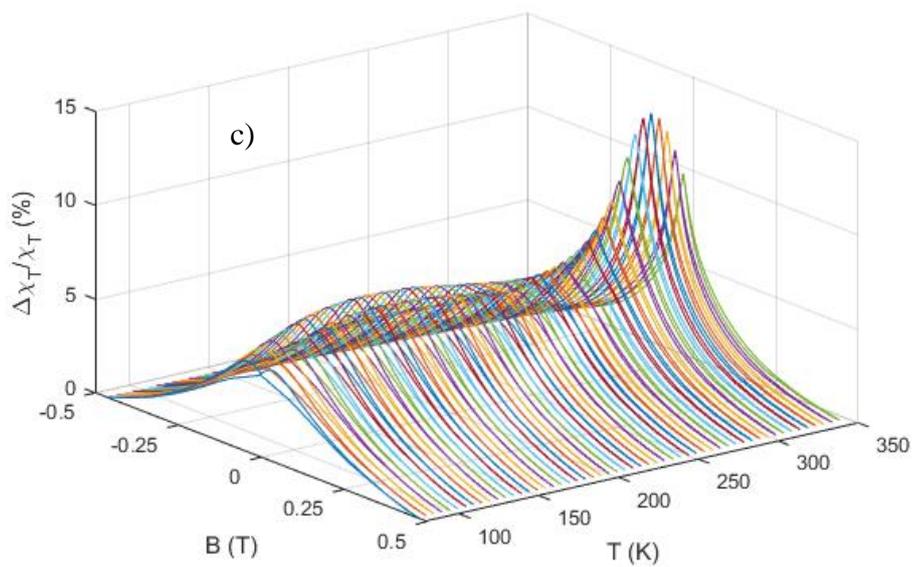

Figure 4

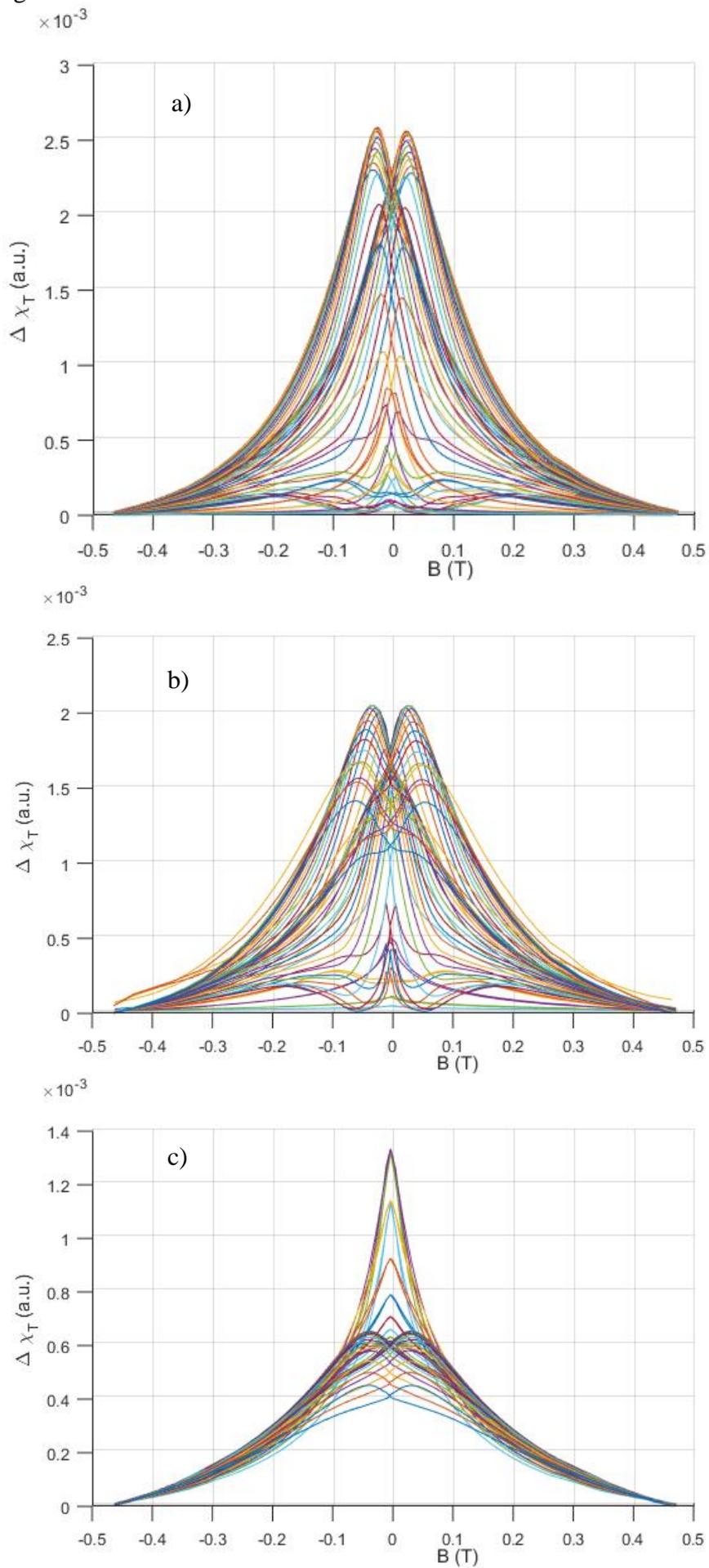

Figure 5

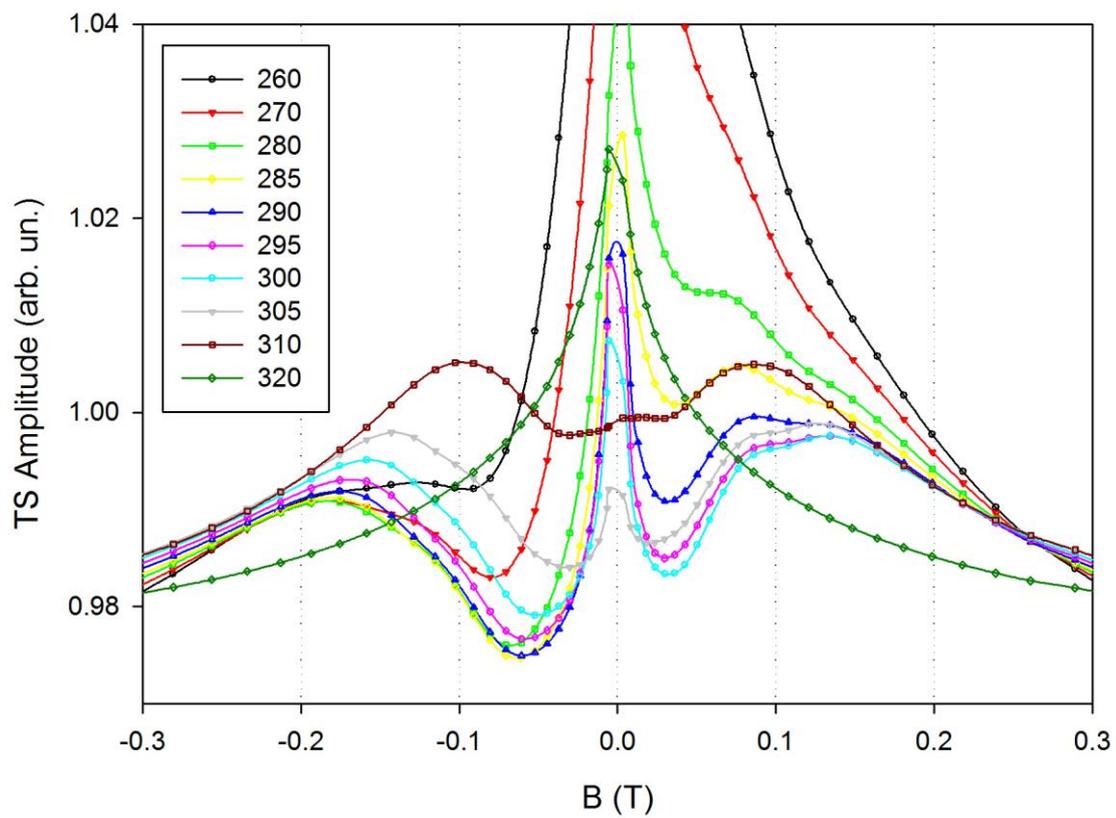